\documentclass[twocolumn,showpacs,preprintnumbers,amsmath,amssymb]{revtex4}

\begin{document}
\draft

\title{Soliton like wave packets in quantum wires}

\author{P. Singha Deo}
\address{Unit for Nano Science and Technology,
S. N. Bose National Centre for Basic Sciences, JD Block,
Sector III, Salt Lake City, Kolkata 98, India.}
\date{\today}

\begin{abstract}
At a Fano resonance in a quantum wire there is strong quantum
mechanical back-scattering. When identical wave packets are incident
along all possible modes of incidence, 
each wave packet is strongly scattered.
The scattered wave packets compensate each other in such a way
that the outgoing wave packets are similar to the incoming
wave packets. This is as if the wave packets are not scattered
and not dispersed. This typically happens for the kink-antikink
solution of the Sine-Gordon model. As a result of such
non-dispersive behavior, the derivation of
semi-classical formulas like the Friedel sum rule and the
Wigner delay time
are exact at Fano resonance. 
For a single channel quantum wire this is true for any
potential that exhibit a Fano resonance.
For a multichannel quantum wire we give an easy prescription
to check for a given potential, if this is true.
We also show that
validity of the Friedel sum rule may or may not be related to
the conservation of charge. If there are evanescent modes
then even when charge is conserved, Friedel sum rule may break down
away from the Fano resonances.

\end{abstract}

%\pacs{PACS: }

\maketitle

\section{Introduction}

The Landauer-Buttiker approach to mesoscopic physics is rather novel.
One of the great success of this approach is the Landauer-Buttiker 
conductance formula.
To understand this approach further and the key to generalize
this approach is the Friedel sum rule (FSR). It has been
the subject of much study recently. While exact proofs
can be given for bulk samples \cite{koh,oku,das}, in low dimensional
systems some attempts to derive it ignore the effects of 
the leads \cite{sou,avi}. 
Buttiker and his co-workers, emphasize the effects
of the leads and find a correction term to the FSR \cite{gas,yey}.
They state that leads can result in non-conservation of charge
in quantum regimes and in such regimes FSR will break down. 
When the system is
in the WKB limit, then charge is conserved and FSR works very well.
Recent explicit calculations \cite{swa} for an impurity in a quantum
wire contradicted this result. 
A single attractive impurity in a quantum wire can produce many
resonant states that can all be classified as Fano resonances \cite{ssc,
bag}. Such an impurity in a quantum wire has attracted
many theoretical investigation \cite{boe}.
Ref. \cite{swa} finds that the Friedel
sum rule is exact at the Fano resonance which is a pure quantum interference
phenomenon (and not a WKB regime), 
and very bad in the regimes away from the
Fano resonance (that are in the WKB regime). 
Ref. \cite{lar} shows that other semi-classical formulas
like the Wigner delay time (WDT) also become exact at the
Fano resonance. 
An analysis of charge conservation and the origin
of semiclassical behavior in a quantum regime is
missing in Refs. \cite{swa,lar}.
In this work we show that there is no connection
between charge conservation and validity of FSR
in the sense that FSR can be violated even when charge
is conserved.
We shall also show that although such an impurity in
a quantum wire  give strong back scattering that is
quantum mechanical in nature, such scattering do not disperse
a wave packet. We shall also show that this
explains why semi-classical theories are exact
in a purely quantum mechanical regime.

When one considers transport in mesoscopic systems then one typically
considers a system as shown in Fig 1.
The system between the points A and B is a grand canonical system
coupled to reservoirs. The way we study grand canonical systems
in text books is that the reservoir Hamiltonian and the system
Hamiltonian can be decoupled. This allows one to construct a 
grand canonical partition function.
But mesoscopic samples are so small
that the actual modeling of the coupling to the reservoirs is necessary
\cite{col,but85,akk,cle}.
The leads (here we show only two leads but there can be many)
are ideal wires that connect the system to the reservoirs.
They inject and absorb electrons and also define the correct boundary
conditions for the system. 
The region between A and B is an elastic scatterer. A particle injected
by reservoir 1 will freely propagate along lead 1 and will be
incident on the scatterer between A and B. The reflected part will
be absorbed by reservoir 1 and the transmitted part will be absorbed
by reservoir 2. The absorbed electrons are completely thermalized inside
the reservoirs and their coherence is destroyed.
Phase shifts are defined with respect to
points A and B and not with respect to $\pm \infty$ \cite{gas}. 
Density of states (DOS) is also the
local density of states (LDOS) integrated between the points A and B
\cite{gas}. The scattering problem is completely defined
with the points A and B \cite{gas} provided the
total charge in the region between A and B (or the integrated
LDOS in the region between A and B) is conserved. 
The region outside that can be parametrized with
chemical potential ($\mu$) and temperature ($T$). 
If $\mu$ and $T$ are the same for the two reservoirs, then we
get an equilibrium situation, and if they are different
then we get non-equilibrium situation. 
All this will become explicit in our model calculation.

\section{the scattering solution}

As a simple realization of such a system (as shown
in Fig. 1) in one dimension (1D) we can consider a
double delta function potential in 1D between $x=-l$ and
$x=l$ (see Fig. 2). The free regions $x<-l$ and $x>l$ are the leads.
For a symmetric scatterer in 1D, the scattering matrix is
\begin{equation}
 S = \left(\begin{array}{cc}
 \displaystyle R &
 \displaystyle T \\
 \displaystyle T &
 \displaystyle R
 \end{array}\right)
\end{equation}
where $R$ is the reflection amplitude and $T$ is the transmission amplitude
of the scatterer.

If we consider a two channel quantum wire with a delta function
potential. The scattering matrix will be 4X4 as shown below.
\begin{equation}
S = \left[ \begin{array}{llcl}
R_{11} & R_{12} & T_{11} & T_{12} \\
R_{21} & R_{22} & T_{21} & T_{22} \\
T_{11} & T_{12} & R_{11} & R_{12} \\
T_{21} & T_{22} & R_{21} & R_{22}
\end{array} \right]
\end{equation}
We are using a notation that $S_{11}=R_{11}$ as it is a
reflection amplitude for an electron incident along the first
transverse mode from the left lead and scattered back
to the first transverse mode in the left lead. 
Similarly, $S_{12}=R_{12}$ as it is a
reflection amplitude for an electron incident along the first
transverse mode from the left lead and scattered back
to the second transverse mode in the left lead. 
Similarly, $S_{13}=T_{11}$ as it is a
transmission amplitude for an electron incident along the first
transverse mode from the left lead and scattered forward
to the first transverse mode in the right lead. 
One can easily understand the rest.
One can solve the scattering problem to find that
for $\alpha$ and $\beta$ taking values 1 or 2,
\begin{equation}
R_{\alpha \beta}=-{i\Gamma_{\alpha \beta} \over 2d \sqrt{k_\alpha k_\beta}}
\end{equation}
If $\alpha \ne \beta$ then
\begin{equation}
T_{\alpha \beta}=-{i\Gamma_{\alpha \beta} \over 2d \sqrt{k_\alpha k_\beta}}
\end{equation}
If $\alpha = \beta$ then
\begin{equation}
T_{\alpha \alpha}=1+ R_{\alpha \alpha}
\end{equation}
Here
\begin{equation}
\Gamma_{\alpha \beta}= {2m \gamma \over \hbar^2} Sin[{\alpha \pi \over w}
(y_j+w/2)]  Sin[{\beta \pi \over w} (y_j+w/2)]
\end{equation}
\begin{equation}
d=1+\Sigma_\nu {\Gamma_{\nu \nu} \over 2 \kappa_\nu} +i\Sigma_{\alpha}
{\Gamma_{\alpha \alpha} \over 2 k_\alpha}
\end{equation}
\begin{equation}
\Gamma_{\nu \nu}= {2m \gamma \over \hbar^2} Sin[{\nu \pi \over w}
(y_j+w/2)]  Sin[{\nu \pi \over w} (y_j+w/2)]
\end{equation}
$\nu$ can take any integer value greater than 2 (i.e., $\nu$=3,4,5,...).
$\gamma$ is the strength of the delta function potential situated
at $x=0$ and $y=y_j$. $m$ is particle mass and 
$w$ is the width of the quantum wire. $k_1=\sqrt{{2m
\over \hbar^2} (E-{\pi^2 \over w^2})}$ is the wave vector for the
1st propagating channel.
$k_2=\sqrt{{2m
\over \hbar^2} (E-{4\pi^2 \over w^2})}$ is the wave vector for the
2nd propagating channel.
$\kappa_\nu=\sqrt{{2m
\over \hbar^2} ({\nu^2\pi^2 \over w^2}-E)}$ is the wave vector for the
$\nu$th evanescent channel. $E$ is the incident energy.
The $n$th quasi bound state or the Fano resonance
occur at energies that satisfy the following
equation
\begin{equation}
1+\Sigma_{\nu=n}^{\nu=\infty} {\Gamma_{\nu \nu} \over 2 \kappa_\nu} =0
\end{equation}
At such an energy, there will be a large amount of charge
localized around the impurity and decaying away from
the impurity. One can define the points A and B as the cut off
points beyond which the localized charge has decayed to negligible
values. Also in real systems $\nu$ will have some cut off that
can have several physical origins like decoherence or work function
of the quantum wire.
The $\alpha$th injectivity at a point $q \equiv
(x,y,z)$ is due to the incident electron of velocity
$v_\alpha$ (or $-v_\alpha$). It is defined as
\begin{equation}
\sum_{\beta} \rho_{\alpha \beta}(q)=
{1 \over h|v_\alpha|} |\psi^{(\alpha)}(q)|^2
\end{equation}
where, $h$ is Plank's constant, $v_\alpha={\hbar k_\alpha \over m}$, 
$k_\alpha$ is incident
wave vector, $m$ is particle mass, $q$ represents coordinate 
and $\psi^{(\alpha)}(q)$ is quantum mechanical
wave function due to unit current incident in the $\alpha$th channel.
$\rho_{\alpha \beta}(q)$ is known as the partial local density of states
(PLDOS).  For different possible values of incident wave vector,
we get different injectivities. Summing up for all the injectivities
we get the local density of states (LDOS).
Integrating LDOS over entire spatial coordinates we get
DOS. So DOS will be
\begin{equation}
\rho(E)=\sum_{\alpha=1}^M \int_{-\infty} ^\infty {1 \over h|v_\alpha|}
|\psi^{(\alpha)}(q)|^2 dq
\end{equation}
And
\begin{equation}
\rho^{GC}(E)=\sum_{\alpha=1}^M \int_A ^B {1 \over h|v_\alpha|}
|\psi^{(\alpha)}(q)|^2 dq
\end{equation}

Here suffix GC stands for ``grand canonical".
Here $M$ is the total number of incident channels possible.

\section{Friedel sum rule (FSR)}

If the charge in the region between A and B is conserved then
the scattering problem is completely defined with respect to the
points A and B.
FSR suggests that the DOS in Eq. 12 can be calculated from $S$ matrix,
without any knowledge of the $\psi^{(\alpha)}(q)$ as the $S$ matrix
elements can be determined experimentally \cite{kob1,kob2} as
well as theoretically \cite{bay,eng}.

The FSR can be stated as \cite{swa,pra}
\begin{equation}
{d \theta_f \over dE}
\approx \pi[\rho^{GC}(E)-\rho^{GC}_0(E)]
\end{equation}
where
\begin{equation}
\theta_f={1 \over 2i} log(Det[S])
\end{equation}
$S$ is the scattering matrix of a system and
$E$ is incident electron energy.
$\rho^{GC}(E)$ is integrated LDOS of a system in presence of
scatterer as defined in Eq. 12 and $\rho^{GC}_0(E)$ is integrated
LDOS of the same system
in absence of scatterer, which naturally requires
that impurity scattering conserves the total number of states
in the system or the total charge in the system (or
else $\rho$ need not be related to $\rho_0$ at all).
In Eq. 13 we have used an approximate equality as there will
be a correction term which we will discuss later.
The beauty of Eq. 13 is its universality. At any resonance (or
quasi bound state) $[\rho^{GC}(E)-\rho^{GC}_0(E)]$ change by unity and hence
$\theta_f$ will change by $\pi$. Moreover, $d\theta_f \over dE$
can be determined
from asymptotic wave function ($x \rightarrow \infty$)
and so one can completely
avoid integrating the LDOS to find the DOS.

The purpose of this section is to explain the discrepancy observed
in Ref. \cite{swa} about the FSR. Namely the FSR becomes exact in a purely
quantum regime like the Fano regime and bad away from the Fano regime.
According to our previous understanding, it should have been the
opposite. 
Such an explanation requires a detailed analysis of charge conservation
and quantum behavior as follows.
A physical origin of such a behavior will become clear
in the next section.

To understand where FSR may go wrong, we first inspect a derivation
of the FSR \cite{pra}. We present it for 1D as the steps can be
repeated for Q1D. Suppose there is an extended potential $V(x)$ 
(to be dimensionally correct in the subsequent analysis,
$V(x) \equiv eV(x)$, where $e$ is electronic charge). Assuming that
$S_{\alpha, \beta}(E,V(x))$ is analytic, we can make an
expansion as
$$S_{\alpha, \beta}(E,V(x)+\delta V(x))=S_{\alpha,\beta}(E,V(x)) +$$
\begin{equation}
\int_{-\infty}^{\infty} 
dx' [{\partial S_{\alpha, \beta}(E,V(x')) \over \partial V(x')} \delta V(x')]  + .....
\end{equation}
Essentially this means breaking up the increment $\delta V(x)$
(although an infinitesimal perturbation, it is an extended potential)
into many local increments $\delta V(x')$ and integrating the
effect of all these local increments. $\delta V(x')$ is therefore
a delta function potential at $x'$.
Now without any loss of generality, we can say that $\delta V(x) = V_0$
for all $x$. In other words $\delta V(x)$ is a constant potential. 
Since $\delta V(x) = V_0$ for all $x$, the local perturbation 
$\delta V(x')$ is also equal to $V_0$, numerically. One has to
remember that the two perturbations $\delta V(x) = V_0$ and
$\delta V(x') = V_0$ are actually different. One of them is a
global perturbation or an extended perturbation while the latter is
a local perturbation. However, for 
$V_0 \rightarrow 0$, one can neglect this difference between them
to write
$${S_{\alpha, \beta}(E,V(x)+ V_0)-S_{\alpha,\beta}(E,V(x))
\over V_0} \approx $$
\begin{equation}
\int_{-\infty}^{\infty} dx' [{\partial S_{\alpha, 
\beta}(E,V(x')) \over \partial V(x')} ]
\end{equation}
Note that now we have an approximate equality and this can be
further justified by explicit calculations as shown below.

Now one may propose that instead of increasing the potential everywhere
by an infinitesimal amount $V_0$, one may keep the potential constant
and instead decrease the incident energy by $\Delta E=V_0$. Thus
$${S_{\alpha, \beta}(E-\Delta E,V(x))-S_{\alpha,\beta}(E,V(x))
\over -\Delta E} = $$
$${S_{\alpha, \beta}(E,V(x)+\delta V(x))-S_{\alpha,\beta}(E,V(x))
\over V_0} \approx $$
\begin{equation}
\int_{-\infty}^{\infty} 
dx' [{\partial S_{\alpha, \beta}(E,V(x')) \over \partial V(x')} ]
\end{equation}

One can prove that \cite{pra}
\begin{equation}
-{1 \over 4 \pi i}  
(S^\dagger_{\alpha \beta} {\partial
S_{\alpha \beta} \over \partial V(x')} - HC) = \rho_{\alpha \beta}(x)
\end{equation}
where $\rho_{\alpha \beta}$ is the PLDOS. PDOS is therefore
$\rho'_{\alpha \beta}(E)=\int_{-\infty} ^\infty \rho_{\alpha \beta} (x) dx$.
One can take any potential in 1D and check that this equation is
exact as done in Ref. \cite{gas}.
Therefore, from Eqs. 17 and 18, 
\begin{equation}
{1 \over 4 \pi i} (S^\dagger_{\alpha \beta} {
dS_{\alpha \beta} \over dE} - HC) \approx \rho'_{\alpha \beta}(E)
\end{equation}
This on summing over $\alpha$ and $\beta$ and
further simplification gives
\begin{equation}
{1 \over 2  i }{d \over dE} log(Det[S]) \approx \pi[\rho(E) - \rho_0(E)]
\end{equation}
Thus we have derived FSR.

Replacing $\int dx' {\partial S_{\alpha,\beta} 
\over \partial V(x')}$ by 
-$\partial S_{\alpha,\beta} \over \partial E$ is an approximation.
Thus, $d\theta_f \over dE$ is not exactly equal to $\pi 
[\rho(E) - \rho_0(E)]$
and so naturally one can expect that $d\theta_f \over dE$
is also not exactly equal to $[\rho^{GC}(E) - \rho^{GC}_0(E)]$. 
In fact \cite{yey},
\begin{equation}
{d\theta_f \over dE}= \pi[\rho^{GC}(E) - \rho^{GC}_0(E)] 
- Im(R_{LL} +R_{RR})/4E
\end{equation}
We have used suffixes `$LL$' and `$RR$', instead of
$\alpha$ and $\beta$. The reasons are obvious as
$R_{LL}$ is for the electrons incident from the left and 
reflected back to the left lead, while $R_{RR}$ is for the
electrons incident from the right and reflected
back to the right lead.
One may consider Eq. 21 as a new FSR but the correction term $[Im(R_{LL}+
R_{RR}) /4E]$ is not
very universal. It can be different for different kinds of resonances.
Secondly, in quasi 1D we will see that this correction term will
also depend on internal details of the potential and can vary
from sample to sample.

So the correction term is $Im(R_{LL} +R_{RR})/4E$. Ref. \cite{yey} and others
assume that this term is due to the non-conservation of charge in
the system.  They assume
(see Eqs. 11 and 12 in Ref \cite{yey}) that this term can be related
to self energy due to the escape probability of an electron
in to the leads. So according to \cite{yey}, in quantum regimes, this 
term can be large.
An essential component
of this work is to establish that this correction term is not due
to non-conservation of charge. 
Although in 1D, 2D and 3D the correction term is large when the
escape probability to the leads is large (that is charge
is not conserved in the grand canonical system)
and vice versa, this is not true in Q1D.
We show below that the correction term
can be large in Q1D even when charge is conserved and
also the correction term can be zero in the Fano regime which
is a quantum regime.

It is shown in the appendix that
$$[(\rho(E) - \rho_0(E)] -[\rho^{GC}(E) - \rho^{GC}_0(E)]=$$
\begin{equation}
{sin[2kl] \over k}(R_{LL}+R_{RR}) + {cos[2kl] \over k}(iR_{LL} - iR_{RR}) 
\end{equation}
This has two implications. First is that since $\pi [\rho^{GC}(E) - 
\rho^{GC}_0(E)] - Im(R_{LL} +R_{RR})/4E
\ne \pi [(\rho(E) - \rho_0(E)]$,
it follows from Eq. 21 that
\begin{equation}
{d\theta_f \over dE} \ne \pi[\rho(E) - \rho_0(E)]
\end{equation}
It can only be an approximate equality as shown before.
The second implication is that the correction term
$- Im(R_{LL} +R_{RR})/4E$ is not due to the lack
of charge conservation.
This is explained below.
When we integrate over all energy then we get that
the RHS of Eq. 22 goes as $\delta(k)$.
The global charge has to be conserved, implying
$\int_{-\infty}^{\infty} dE [(\rho(E) - \rho_0(E)]=0$.
Hence from Eq. 22,
$\int_{-\infty}^{\infty} dE [(\rho^{(GC)}(E) - \rho^{(GC)}_0(E)]$
goes as $\delta(k)$.
Since only positive energy states are propagating states
that we are interested in,
one can always take the integration over $E$ in the positive energy
regime instead of taking it from $-\infty$ to $\infty$.
As $k=0$ is a non-propagating state, in the propagating regime
$\int_{\epsilon}^{\infty} dE [(\rho^{(GC)}(E) - \rho^{(GC)}_0(E)]$=0.
So charge is conserved in the grand canonical system.
So the correction term in Eq. 21 is arising due to the error involved in the
substitution in Eq. 17 and has nothing to do with charge conservation.
It is just an error due to an approximation in the algebra.

Although in the appendix we have considered
a 1D system, all the steps can be repeated for a single channel Q1D system.
Only the
expressions for $R_{LL}$ and $R_{RR}$ will be different and
$k=\sqrt{{2m_e \over \hbar^2}(E-E_1)}$.
So for a single channel quantum wire, 
\begin{equation}
{d \theta_f \over dE} = \pi[\rho^{GC}(E)-\rho_0^{GC}(E)] -
Im{R_{LL}+R_{RR} \over 4(E-E_1)} 
\end{equation}
First of all note the presence of sample specific parameter $E_1$
in the correction term that was ignored in Ref. \cite{yey}. 
This equation is the same whether evanescent modes are present or
not present. However, only the expressions for $R_{LL}$ and 
$R_{RR}$ changes
completely in presence and absence of evanescent modes.
From Eq. 3 
\begin{equation}
-Im[R_{LL}+R_{RR}]={{\Gamma_{11} \over k_1} (1+\sum_{n>1} {\Gamma_{\nu \nu}
\over 2 \kappa_\nu}) \over (1+\sum_{n>1} {\Gamma_{\nu \nu}\over
2 \kappa_\nu})^2 + ({\Gamma_{11} \over 2k_1})^2}
\end{equation}
For a delta function potential in 1D,
\begin{equation}
-Im[R_{LL}+R_{RR}]={{\Gamma_{1D} \over k_1} 
\over 1+
({\Gamma_{1D} \over 2k_1})^2}
\end{equation}
where $\Gamma_{1D} = {2m\gamma \over \hbar^2}$.
In comparison with the 1D case, the only difference in quasi one
dimension (Q1D) (compare Eqs. 25 and 26) is the term 
\begin{equation}
\sum_{\nu>1}\frac{\Gamma_{\nu \nu}}{2\kappa_\nu}
\end{equation}
If we remove this term then the correction term is negligible
for $k_1>\Gamma_{11}$ which is the semi-classical regime.
Complications in Q1D arise because of the series term
$\sum_{n>1}\frac{\Gamma_{\nu \nu}}{2\kappa_\nu}$. 
Even for $k_1<\Gamma_{11}$, $(1+\sum_{n>1} {\Gamma_{\nu \nu}\over
2 \kappa_\nu})$ can become zero and then the correction term can become 0
in a purely quantum regime. 
At the Fano resonance this is exactly what happens, i.e.,
RHS of Eq. 25 becomes 0 precisely due to the fact that
$(1+\sum_{n>1} {\Gamma_{\nu \nu}\over
2 \kappa_\nu})$=0 at the Fano resonance (see Eq. 9).
Also note that
although each term in the
series decreases with energy, the sum does not decrease easily
as the series is a divergent series. It goes as $log[N]$ where
$N$ is the total number of terms in the series or the total
number of evanescent modes \cite{boe}.
One can make the transverse width $w \rightarrow \infty$
to create an infinite number of evanescent modes and then one can
see from Eq. 25 that the correction term goes to zero implying that
FSR is exact in 2D.
In real quantum wires, we have to truncate the series at some value $N$.
For any arbitrary number of evanescent modes,
the correction term can be as large as
$d \theta_f \over dE$ or $\pi[\rho^{(GC)}(E)-\rho^{(GC)}_0(E)]$,
making the two qualitatively and quantitatively different,
except in a narrow energy regime close to the upper band edge.
At the upper band edge $\sum_{n>1}\frac{\Gamma_{\nu \nu}}{2\kappa_{\nu}}$
diverge as the first term in it (i.e., $\Gamma_{22} \over 2k_2$) 
diverges and hence RHS of Eq. 25
becomes 0.

\section{wigner delay time (wdt)}

The fact that FSR becomes exact at the Fano resonance is very
counterintuitive. FSR is similar to WDT and so it was also
checked that WDT at the Fano resonance becomes exact \cite{lar}. 
The similarity between WDT and FSR can be seen from Eqs. 19 and 20.
$$\sum_{\alpha \beta} {1 \over 4 \pi i} [S^\dagger_{\alpha \beta}
{dS_{\alpha \beta} \over dE} - HC] =$$ 
\begin{equation}
\sum_{\alpha \beta} {1 \over 2 \pi}[|S_{\alpha \beta}|^2 {d\over dE} 
arg(S_{\alpha \beta})]
=\sum_{\alpha \beta} \int_{-\infty}^\infty \rho_{\alpha \beta}(x) dx
\end{equation}
$\hbar {d \over dE} arg(S_{\alpha \beta})$ is the WDT for particles
transmitted from the $\alpha$th channel to the $\beta$th channel
and there are $|S_{\alpha \beta}|^2$ of such particles. One can
choose $\hbar=1$.
Here $arg(S_{\alpha \beta})=Arctan[{Im[S_{\alpha \beta}] \over 
Re[S_{\alpha \beta}}]$.
We have also seen that the LHS in Eq. 28 is the semi-classical
limit of the LHS of Eq. 18 integrated over $x'$. So in the semi-classical
limit, WDT times the number of particles involved 
gives the PDOS. 
It was shown in Ref. \cite{swa}, that in the
Fano regime also the WDT (${1 \over 2 \pi} |S_{\alpha \beta}|^2 {d\over dE}
arg(S_{\alpha \beta})$) gives the PDOS ($ \int_{-\infty}^\infty 
\rho_{\alpha \beta}(x) dx$) exactly,
in spite the fact that Fano resonance is a quantum
phenomenon. This happens for single channel quantum
wires as well as for multi channel quantum wires.
Another way to see that the WDT is semi-classical is that
its derivation is based on non-dispersive wave-packets.
Below we show how non-dispersive wave-packets are
realized in the quantum regime of Fano resonance and as a
result WDT becomes exact (that is WDT gives
the PDOS correctly).

We start by presenting a derivation of the WDT based on
non-dispersive wave-packets.
Let us consider an incident Gaussian wave packet in 1D
representing an ensemble of non-interacting particles.
$a(k)$ is the weight of the $k$th Fourier component
in the incident Gaussian wave packet.
\begin{equation}
\psi_{in}(x,t)=\int_{-\infty}^{\infty} a(k) exp[ikx - iwt] dk
\end{equation}
After the wave packet traverses a distance $L$, its form will be
\begin{equation}
\psi_{tr}(x,t)=\int_{-\infty}^{\infty} a(k) T(k) exp[ik(x+L) - 
iw(t+t_0+\Delta t)] dk
\end{equation}
Here, $T(k)$ is the transmission amplitude of the potential in
the region of length $L$.
$t_0$ is the time that the wave packet would have taken if the potential
was absent. $t_0+\Delta t$ is the time that the wave packet takes in presence
of the potential. 
If we go to the semi-classical limit
then we should get close to classical behavior
that implies $\psi_{tr}(x,t)$ is also a Gaussian wave-packet
like $\psi_{in}(x,t)$. From this one can derive WDT.
Normally, $T(k)$ is complex and
energy dependent. 
This is the essential cause of dispersion. 
The weight of the $k$th component in the transmitted
wave packet are $a(k)T(k)$ 
and hence $\psi_{tr}$
is no longer a Gaussian wave packet.
If $T(k)$ is a real number,
then the dispersion will be like a free particle as $k$ and $w$ 
in $\psi_{tr}$ are
identical to that of a free particle ($w={\hbar k^2 \over 2m}$).
One simple example where this happens is when the incident energy is much 
smaller than the potential height, wherein one gets
$T(k) \rightarrow$0 and $R(k) \rightarrow$-1. 
In this case $R(k)$ is real.
One finds the WDT 
\begin{equation}
\Delta t = \hbar {d \over dE} arg(R) = {d \over dw} arg(R)
\end{equation}
and it correctly gives the PDOS (that is ${1 \over 2 \pi}
|R|^2 {d \over dE} arg(R) =
\int \rho_{\alpha \beta}(x) dx$).

This explains why FSR is
exact in case of single channel Fano resonance where the
particle is completely reflected back due to an effective
potential that is infinite. At the single
channel Fano resonance
$R(k)=-1$ and WDT give the correct PDOS. 
This also shows that the correctness of WDT and hence FSR at
Fano resonance is always true in single channel quantum wires.
It requires the presence of a transmission zero and that is always
there for all potentials that support a Fano resonance.
But the correctness of FSR or WDT does not only occur
in case of single channel quantum wires where $R(k)$=-1 as
in semi-classical limit, but it also happens in multi channel
quantum wires where 
$|R_{11}(k)|\ne$0 and $|T_{11}(k)|\ne$0. So how
for such a system WDT or FSR remain exact?

In order to show how one can get non-dispersive wave packets in
the presence of quantum scattering, we take clue from
the kink-antikink solution of the Sine-Gordon equation.
Suppose we have a delta function potential in a 2 channel
quantum wire. 
Let us have four identical Gaussian wave
packets incident on it along all possible channels.
That means two will be incident from the left and two
will be incident from the right. Among the two that are incident from the
left, one will be in the first channel or in the fundamental
transverse mode and one will be in the 2nd channel that is the first
excited transverse mode. Similarly for the two that are incident from
the right. All these
wave packets are scattered at the same time and we call this time
$t$. After scattering, the resultant wave packet on the
right in the fundamental mode
(say) and moving away from the potential and at a distance
$L$ from the delta function potential will be
$$\psi_{tr}^{QW}=
\int a_1(k_1) T_{11}(k_1) exp[ik_1(x + L) - iw(t+t_0+
\Delta t_{T_{11}}]dk_1$$ 
$$+\int a_2(k_2) T_{21}(k_2) exp[ik_1(x + L) - 
iw(t+t_0+ \Delta t_{T_{21}}]dk_2$$ 
$$+\int a_1(k_1) R_{11}(k_1) exp[ik_1(x + L) - 
iw(t+t_0+ \Delta t_{R_{11}}]dk_1$$ 
\begin{equation}
+\int a_2(k_2) R_{21}(k_2) exp[ik_1(x + L) - iw(t+t_0+ \Delta t_{R_{11}}]dk_2 
\end{equation}
Here $t_0+ \Delta t_{T_{11}}$ is for example the time taken
by a particle in going from first channel in the
left lead to the first channel on the right lead, and so on.
One has to start with an infinitesimal potential so that with a small
probability a particle goes from channel 2 on left to channel 1 on
the right with an infinitesimal $\Delta t_{T_{21}}$. 
And then by increasing the potential to its actual
value, one can get actual $\Delta t_{T_{21}}$ etc.
It is easy to show that $a_1(k_1)=a_1(-k_1)$ and $a_2(k_2)=a_2(-k_2)$.
So FSR as well as WDT will be correct if $\psi_{tr}^{QW}$ is also a 
Gaussian wave packet. One way to get that is if $T_{11}(k)$,
$T_{21}(k)$, $R_{21}(k)$ and $R_{11}(k)$ are simultaneously real.
Because then the weight of the $k$th component is a real number
times $a_i(k)$ and further those real numbers are complimentary to each
other and also summed. That is if $T_{11}(k)$ increases then
$T_{21}(k)$ decreases and the first two terms in Eq. 32 compensate
each other and so on.
One can also show that $T_{21}=R_{21}$, and $arg(T_{21})
= arg(R_{21}) =  arg(R_{11})$.
In the following figure we show that $T_{11}$, $T_{21}$,
$R_{21}$ and $R_{11}$ are simultaneously real at
the Fano resonance. Since they are real, their squares add up to
make 1. So they are also complementary to each other and compensate
each other. If $T_{11}$ is small the $R_{11}$ is large and so on.
Actually, all the phase shifts vary strongly with energy as is expected
in a quantum regime, but the variations are around 0
and becoming exactly 0 at the Fano resonance. 

One can check the outgoing wave packets in the other channels also.
They all show similar behavior at the Fano resonance. 
Although,
the individual wave packets get strongly scattered, the four
scatterings compensate each other in such a way that the outgoing
waves are similar to the incoming waves. 
So the derivation of WDT
holds good and so naturally the WDT also holds good. And then
summing over all the particles making the wave packet, one
naturally gets that FSR holds good. This provides a physical
picture that helps us to understand why semi-classical formulas
based on un-dispersed wave packets
hold good in an extreme quantum regime. 
Semi-classical
formulas are always much simpler and easy to understand as it
has classical analogies.

\section{conclusion}

For larger systems, that is when the sample size is larger than
the inelastic mean free path, it has been argued that the scattering
matrix approach do not take into account the conservation of
charge \cite{mpd}. 
FSR can break down due to non-conservation of charge \cite{ll}.
In this work we show that even for mesoscopic systems,
that is when sample size is smaller than inelastic mean free path,
although charge is conserved, the scattering matrix approach
does not give the DOS exactly.
In a quantum wire,
the correction term due to the evanescent modes is quite
complicated and it is not possible to make any general
statement about it like
correction term is negligible in semi-classical
regime and large in quantum regimes.
Quite counter intuitively, the correction term in Eq. 25,
becomes 0 at the Fano resonance
as a result of which the
FSR becomes exact. We do not know of any system where this
correction term can become exactly 0. 
We have shown that in single channel quantum wire, this is true
for all potential that exhibit a Fano resonance as it only
requires the presence of a transmission zero.
We have also taken a scatterer in a multi channel quantum wire that has
Fano resonance, wherein all the $S_{\alpha \beta}$s are
non zero and also strongly energy dependent.
But the
correction term is once again exactly 0 making the FSR
exact at the Fano resonance.
We provide a physical understanding of this based on
non-dispersive wave packets that are crucial for the derivation
of semi-classical formulas like FSR and WDT. 
This gives us a general prescription to check for a 
given Fano resonance in a multi-channel quantum wire,
if semi-classical formulas will be exact or not.
Although, the
quantum mechanical scattering can strongly disperse the different
partial waves, the resultant of all possible partial waves
in the Hilbert space and their scattering compensate each
other in such a way that the resultant wave-packet is un-dispersed.

The advantage of using FSR to know the DOS of a system has certain
advantages. It makes it un-necessary to find the
local wave-functions inside a scatterer and also removes the
problem of integrating the LDOS to find the DOS. Also FSR is expected
to work in presence of electron-electron interactions \cite{lan}.
An easy way to see this is to consider the Kohn-Sham theorem \cite{kon},
which essentially means that an electron passing through an interacting
system, actually passes through a one body effective potential that
accounts for exchange and correlation effects exactly.

\section{acknowledgment}
The author is grateful to ICTP, Italy for support where a part
of this work was done.

\section{appendix}

Let us calculate the DOS $\rho(E)$ for the system in Fig. 2.
We first consider the electron incident from the left (as shown in Fig. 2),
with incident wave vector $k$. The PDOS in this case is
$$\rho^{(1)}(E) = {1 \over h|v|} \int ^l_{-l} \mid a e^{ikx} + b
e^{-ikx} \mid ^2 dx$$ 
\begin{equation}
+ \int^{-l}_{-\infty} |e^{ikx} + R e^{-ikx}|^2 dx
+ \int_l^{\infty}|Te^{ikx}|^2dx
\end{equation}
$T$ is the same whether incident from left or incident from right.
We next consider the electron incident from the right,
with incident wave vector $-k$. The PDOS in this case is
$$\rho^{(2)}(E) = {1 \over h|v|} \int ^l_{-l} \mid a e^{-ikx} + b
e^{ikx} \mid ^2 dx$$ 
\begin{equation}
+ \int_{l}^{\infty} |e^{-ikx} + R e^{ikx}|^2 dx
+ \int^{-l}_{-\infty}|Te^{-ikx}|^2dx
\end{equation}
Therefore DOS is given by 
$$\rho(E)=\rho^{(1)}(E) + \rho^{(2)}(E)=
{1 \over hv}[ 2 \int_{-\infty}^{\infty} dx
+ 2 \rho'$$
$$+ R \int_{-\infty}^{-l} cos[2kx] dx
+ iR \int_{-\infty}^{-l} sin[2kx] dx$$
$$ + R \int^{\infty}_{l} cos[2kx] dx
+ iR \int^{\infty}_{l} sin[2kx] dx$$
$$+ R \int_{-\infty}^{-l} cos[2kx] dx
- iR \int_{-\infty}^{-l} sin[2kx] dx$$
\begin{equation}
+ R \int^{\infty}_{l} cos[2kx] dx
- iR \int^{\infty}_{l} sin[2kx] dx
\end{equation}
where
\begin{equation}
\rho'=\int_{-l}^l |a e^{ikx} + b e^{-ikx}|^2 dx
- 2 \int_{-l}^{l} dx ={hv \over 2 \pi}[(\rho^{GC}(E) - \rho^{GC}_0(E)]
\end{equation}
The indefinite integrals on $sin[x]$ and $cos[x]$ can be done by breaking
them up in exponential functions to give
$$\rho(E)={1 \over hv}[2 \int_{-\infty}^{\infty} dx $$
\begin{equation}
+ 2 \rho' - {sin[2kl] \over k}(R+R) +{cos[2kl] \over k}
(iR - iR)]
\end{equation}
Thus we have proved that 
$$\rho(E)-\rho_0(E)=
\rho^{GC}(E) - \rho^{GC}_0(E)$$
\begin{equation}
- {sin[2kl] \over k}(R+R) +{cos[2kl] \over k}(iR - iR)]
\end{equation}

\centerline{\bf Figure Captions}

\noindent Fig. 1. A grand canonical system, extending from A to B,
connected to two reservoirs on two sides with ideal leads. 

\noindent Fig. 2. A realization of the system shown in Fig. 1 in one
dimensions. 

\noindent Fig. 3. Here G=$1+\sum_{\nu=n}^{\nu=\infty}{\Gamma_{\nu \nu}
\over 2 k_\nu}$, that is 
the LHS of Eq. 9. It is shown by the solid line. When it crosses
the energy-axis, then we get a bound state. $arg(T_{11})$ (dashed curve) 
and $arg(R_{11})$ (dotted curve) become simultaneously 0 at the
bound state or at the Fano resonance.
This implies $T_{11}$, $R_{11}$, $T_{21}$ and $R_{21}$ are simultaneously
real at the Fano resonance.

\end{document}